\documentclass[aip,pop,preprint]{revtex4-1}
\usepackage{graphicx}%
\usepackage{dcolumn}%
\usepackage{bm}%
\usepackage{amsmath}%
\begin{document}
\title{Particle Heating and Energy Partition in Low-$\beta$ Guide Field Reconnection with Kinetic Riemann Simulations}
\date{\today}
\author{Qile Zhang}
\author{J. F. Drake}
\author{M. Swisdak}
\affiliation{University of Maryland, College Park, Maryland, 20742, USA}
\begin{abstract}
Kinetic Riemann simulations have been completed to explore particle
heating during guide field reconnection in the low-$\beta$ environment
of the inner heliosphere and the solar corona. The
reconnection exhaust is bounded by two rotational discontinuities (RD)
and two slow shocks (SS) form within the exhaust as in
magnetohydrodynamic (MHD) models. At the RDs, ions are accelerated by
the magnetic field tension to drive the reconnection outflow as well
as flows in the out-of-plane direction. The out-of-plane flows stream
toward the midplane and meet to drive the SSs. The SSs differ greatly
from those in the MHD model. The turbulence at the shock fronts and
both upstream and downstream is weak so the shocks are laminar and
produce little dissipation. Downstream of the SSs the counterstreaming
ion beams lead to higher density, which leads to a positive potential
between the SSs that acts to confine the downstream electrons to
maintain charge neutrality. The potential accelerates electrons from
upstream of the SSs to downstream region and traps a small fraction
but only modestly increases the downstream electron temperature above
the upstream value. In the low-$\beta$ limit the released magnetic
energy is split between bulk flow and ion heating with little energy
going to electrons. That the model does not produce
strong electron heating nor an energetic electron component suggests that other mechanisms, such as multiple x-line reconnection, are required to explain energetic electron production in large flares. The model can
be tested with the expected data from the Parker Solar Probe.
\end{abstract}

\maketitle

\section{Introduction}
Magnetic reconnection converts magnetic energy into particle kinetic
energy by magnetic field line contraction after a change of magnetic
topology. It drives explosive energetic events in our solar system,
including solar flares, Coronal Mass Ejections (CMEs), and geomagnetic
storms, which can have large impacts on the space weather environment
and even power grids on Earth. However, the conversion process from
the magnetic field to high speed flows, heating and energetic
particles remains only partially understood.

A long lasting puzzle in astrophysics is how particles in the solar
corona are heated through reconnection. While the corona is a
magnetically dominated low-$\beta$ ($\beta$ is the ratio of plasma
thermal pressure to magnetic pressure) environment, the electron
temperature is millions of degrees Kelvin on average and can be even
one to two orders of magnitude hotter in impulsive events such as
solar flares. Reconnection is one of the most promising candidates to
explosively convert magnetic energy into plasma energy, but the
detailed mechanism behind particle heating remains unclear.

How plasma gains energy during reconnection has previously been
analyzed numerically with both fluid and kinetic
descriptions. Sophisticated MHD simulations can employ computational
domains of coronal scales and provide direct comparisons to
observations\cite{Guidoni2010,Guidoni2011}, but do not distinguish
between the heating of electrons and ions, and require
assumptions on particle velocity distributions, isotropy, viscosity
and heat flux without capturing many potentially important kinetic
effects.  Capturing such effects from first principles requires full
particle descriptions. Previous studies with full particle
models are typically localized near the reconnection diffusion
regions\cite{Drake2003,Pritchett2004} or explore outflows from single  \cite{Liu2012,Shay2014,Haggerty2015,Egedal2015,Drake2009,Drake2009a}
or multiple reconnection sites 
\cite{Daughton2011,Dahlin2014,Dahlin2015}.
Due to the computational constraints of conventional kinetic
reconnection simulations, the results are often limited to low
ion-to-electron mass ratio with computational domains that are at most
several hundred $d_i$ in size, where $d_i$ is the ion inertial
length. High mass-ratio and low $\beta$ simulations typically have
smaller computational domains because of the requirement in
particle-in-cell (PIC) models that the Debye length be
resolved. Observations of reconnection in the
magnetosphere\cite{Phan2013,Phan2014} find an empirical linear scaling
for ion and electron heating as a function of the available magnetic
energy per particle, which is consistent with that found in
simulations\cite{Shay2014}. However, these studies only explored the
$\beta$ of order unity regime. The mechanism of electron heating is
under investigation and debate\cite{Haggerty2015,Egedal2015}.

Some of the drawbacks of conventional kinetic reconnection simulations
of a single reconnection outflow can be addressed, in part, by
employing quasi-1D particle-in-cell Riemann simulations so that the
spatial scale along the inflow direction can be dramatically
increased, the upstream plasma $\beta$ can be reduced and the mass
ratio can be increased. This is particularly useful in low $\beta$
systems like the corona since a near-realistic ion-to-electron mass
ratio is necessary to keep the electron thermal speed much greater
than the Alfv\'en speed (as it is in the corona). Riemann simulations
model reconnection outflows in order to study the physics of particle
heating downstream of the ion diffusion region. During reconnection,
the magnetic energy is mostly dissipated downstream of the
reconnection site, where the field lines contract, so it is not
necessary to simulate the reconnection diffusion region in order to
capture the physics of particle heating in a large-scale
system. Instead, the contracting field lines in a Riemann simulation
will uncover the physical processes of particle heating in a single
reconnection outflow. Riemann simulations have been used to explore
the structure of the exhaust but did not investigate particle heating
and in particular the relative heating of electrons and ions. They were based on MHD models\cite{Lin1993},  hybrid simulations\cite{Lin1996,Scholer1998,Cremer1999,Cremer2000} as well
as PIC simulations\cite{Liu2011(a),Liu2011(b)} without a guide field. On the other hand, Riemann simulations do
not address the physics of multi x-line reconection since they presume
that the reconnected magnetic field is uni-directional across the
current layer.

This paper presents investigations of particle heating in low $\beta$
reconnection outflows downstream from a single x-line through PIC
Riemann simulations. Since coronal reconnection typically includes a
guide field, in these simulations the ratio of the guide field to the
reconnecting component of the field is taken to be of the order of or
greater than unity. As in the MHD model we find that there are two
rotational discontinuities (RD) that bound the exhaust and two slow
shocks (SS) that develop within the exhaust. The ions are accelerated at
the RDs and form counterstreaming beams downstream of the
SSs. However, these counterstreaming beams are stable so that
turbulence within the entire exhaust remains weak and as a result SSs
produce little dissipation, an important difference from the MHD
description. Downstream of the SSs the counterstreaming ion
beams produce an increase in their density (by about a factor of
two). A positive potential in the region downstream of the SSs
develops to confine the downstream electrons in this high density
region. The electrons are accelerated by the potential from upstream of the SS
to downstream of the SS and are partly trapped by the potential in the region
between the two SSs. Electron trapping by this potential modestly
increases the downstream electron temperature.

In a series of simulations carried out with increasing upstream
magnetic energy per particle (at fixed upstream temperature) the ion
downstream temperature increases in a linear manner, proportional to
the available magnetic energy, while the electron temperature
plateaus, increasing only modestly from the upstream value. This is
because the electron heating is limited by the amplitude of the
potential across the SSs. A very large potential does not develop
because it would trap too many electrons compared with the modest
increase in ion density and so charge neutrality would be
violated. Thus, neither species undergoes the canonical diffusive shock
acceleration at the SSs since no turbulence scatters particles back
and forth across the shocks. Most of the released magnetic energy goes into
ions driven at the RDs as the bulk reconnection outflow or as the
counterstreaming ion beams in the midplane of the exhuast, which are
not thermalized by the SSs.

The organization of this paper is the following: in Sec.\ 2 the
Riemann simulations are introduced for reconnection modeling; in
Sec.\ 3 the results of simulations are presented and the heating
mechanisms are discussed; in Sec.\ 4 the scaling of electron and ion
heating and energy partition with increasing available magnetic energy
is discussed; and finally the conclusions and implications are in
presented in Sec.\ 5.
\section{Riemann simulations as proxies for reconnection outflows}
\subsection{Riemann simulations}
The magnetic geometry of a Riemann simulation resembles a single
reconnection outflow from the x-line. It reduces the dimension of a 2D
outflow by neglecting the weak dependence on the outflow direction,
thus transforming it into a 1D problem. The time development of the 1D Riemann simulation is a proxy for the time development of the
reconnection exhaust in the frame of the outflow. Since,
in the frame of the outflow, the exhaust expands in width, the results
of a Riemann simulation expand over time as well. In practice, the computational domain of a
Riemann simulation consists of a thin, long, quasi-1D box extended along
the reconnection inflow direction, $y$, with the outflow
direction, $x$, and guide field direction, $z$, short. We take all
boundaries to be periodic. Although our domain contains two current
sheets to achieve periodic boundaries in $y$, we only focus on one
current sheet, as will be described later. The lengths of the two
short dimensions ($x$ and $z$) can be adjusted to include the
wavelengths of the dominant instabilities if they are important. In this way, with
the same computational cost, we can explore the physics of magnetic
energy conversion and particle heating in a large spatial domain with
low-$\beta$ and and relatively high mass ratio. In contrast, with a
conventional reconnection simulation, because the width to length
ratio of the exhaust is around 0.1, it is a challenge to model a
system that is large enough to separate the reconnection exhaust structures
transverse to the outflow direction.

Because the coronal environment typically has low $\beta$, we use a
force free configuration with a guide field where the
initial magnetic field strength and density are constant but there is
magnetic shear at the current sheet. The equilibrium is assumed to be symmetric across each current layer. We also use a small constant
initial $B_y$ (the reconnected component of the magnetic field) to provide the magnetic tension to drive the outflow. The equilibrium is defined as follows:
\begin{subequations}
\label{initallequations}
\begin{eqnarray}
B_x=B_{\text{x,a}}\ \tanh(y/w_0),\label{initequationa}
\\
 B_z=\sqrt{(B_{\text{x,a}}^2+B_{\text{z,a}}^2-B_x^2)},\label{initequationb}
 \\
 B_y=0.1B_{\text{x,a}},
 \\
 n=n_0,\label{initequationc}
 \end{eqnarray}
\end{subequations} 
where the "a" subscript represents asymptotic upstream values. 

We use the particle-in-cell code {\tt p3d} in which the particle
positions and velocities evolve via the Newton-Lorentz equations of
motion \cite{Zeiler2002three‐dimensional}. The electromagnetic fields are advanced in
time by Maxwell's equations.  Magnetic field strengths are normalized
to $B_{\text{x,a}}$, densities to the initial density $n_0$, lengths to
the ion inertial length $d_i=c/\omega_{pi}$ (based on $n_0$), times to
the inverse ion cyclotron frequency $\Omega_{ci}^{-1}$, velocities to
the Alfv\'en speed $C_{Ax}$ based on $B_{\text{x,a}}$ and $n_0$, and
temperatures to $m_iC_{Ax}^2$. The initial conditions used for all the
simulations are shown in Table~\ref{table1}. After a Riemann
simulation starts, the exhaust begins to form and expand in width,
heating ions and electrons within it.

\subsection{Comparison with a reconnection simulation}
\begin{figure*}
\includegraphics[width=\linewidth]{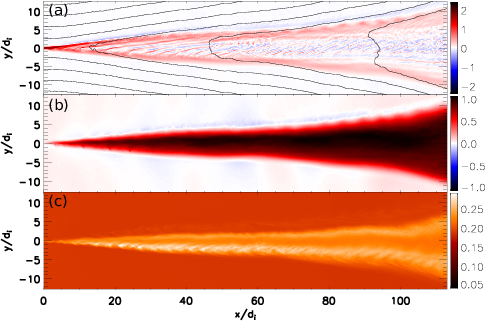}
\caption{\label{fig1} The exhaust of a 2D reconnection simulation (Run 1). (a) $J_z$ with in-plane magnetic field lines overploted, (b) $V_{ix}$, (c) $T_{i\parallel}$.}
\end{figure*}

\begin{figure*}
\includegraphics[width=\linewidth]{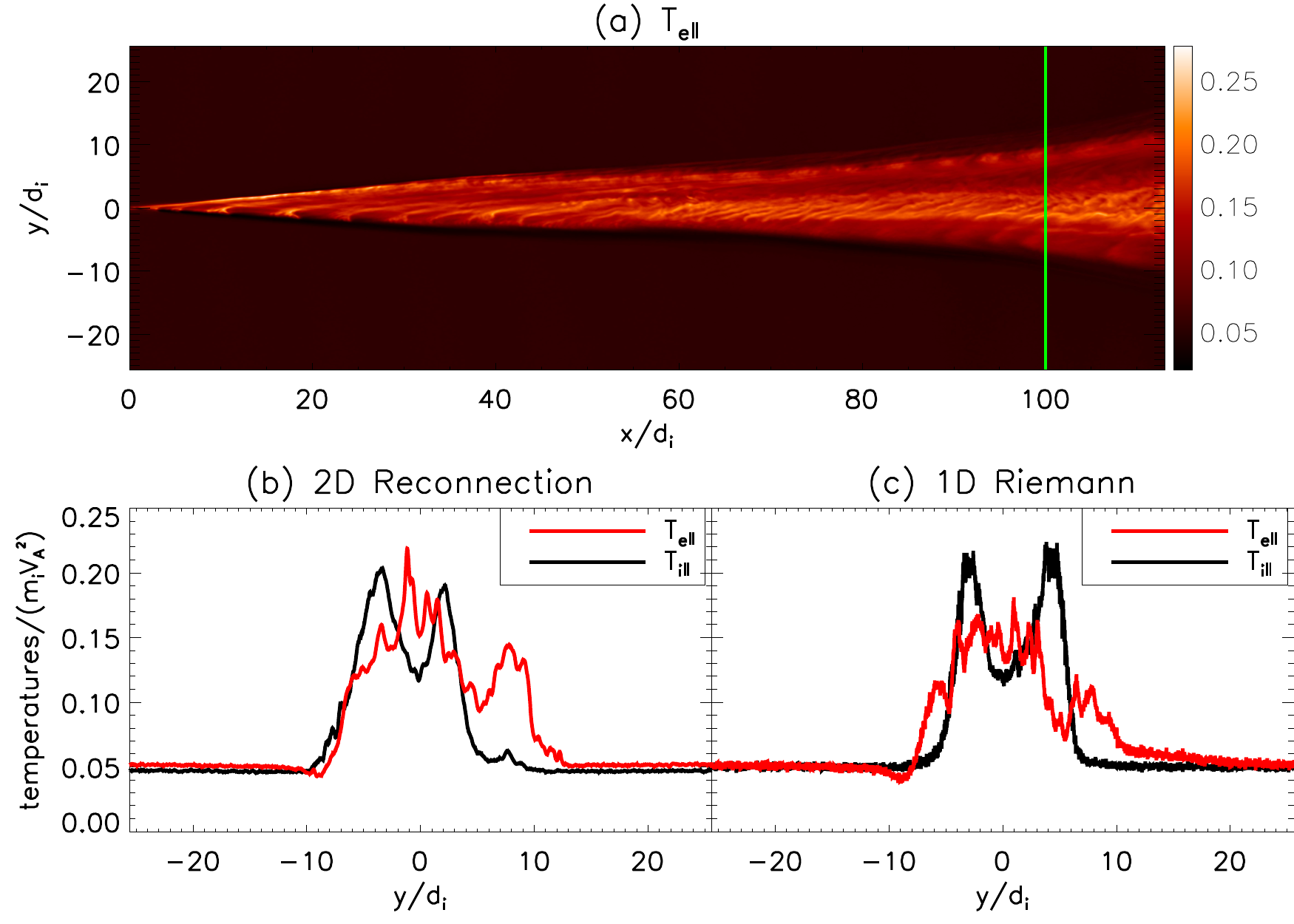}
\caption{\label{fig2} Comparing a 2D reconnection simulation and a 1D Riemann simulation. (a) $T_{e\parallel}$ of the 2D reconnection simulation (Run 1) exhaust, (b) profiles of $T_{e\parallel}$ and $T_{i\parallel}$ taken at the green line of (a), (c) the same profiles from the corresponding 1D Riemann simulation (Run 2).}
\end{figure*}

\begin{figure*}
\includegraphics[width=\linewidth]{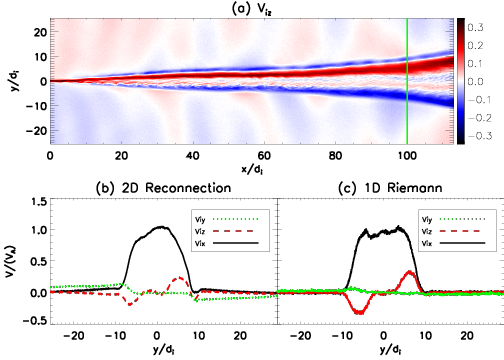}
\caption{\label{fig3} Similar to Fig.2, (a) $V_{iz}$ of the 2D reconnection simulation (Run 1) exhaust, (b) profiles of $V_{ix}$, $V_{iy}$ and $V_{iz}$ taken at the green line of (a), (c) the same profiles from the corresponding 1D Riemann simulation (Run 2).}
\end{figure*}

\begin{figure*}
\includegraphics[width=\linewidth]{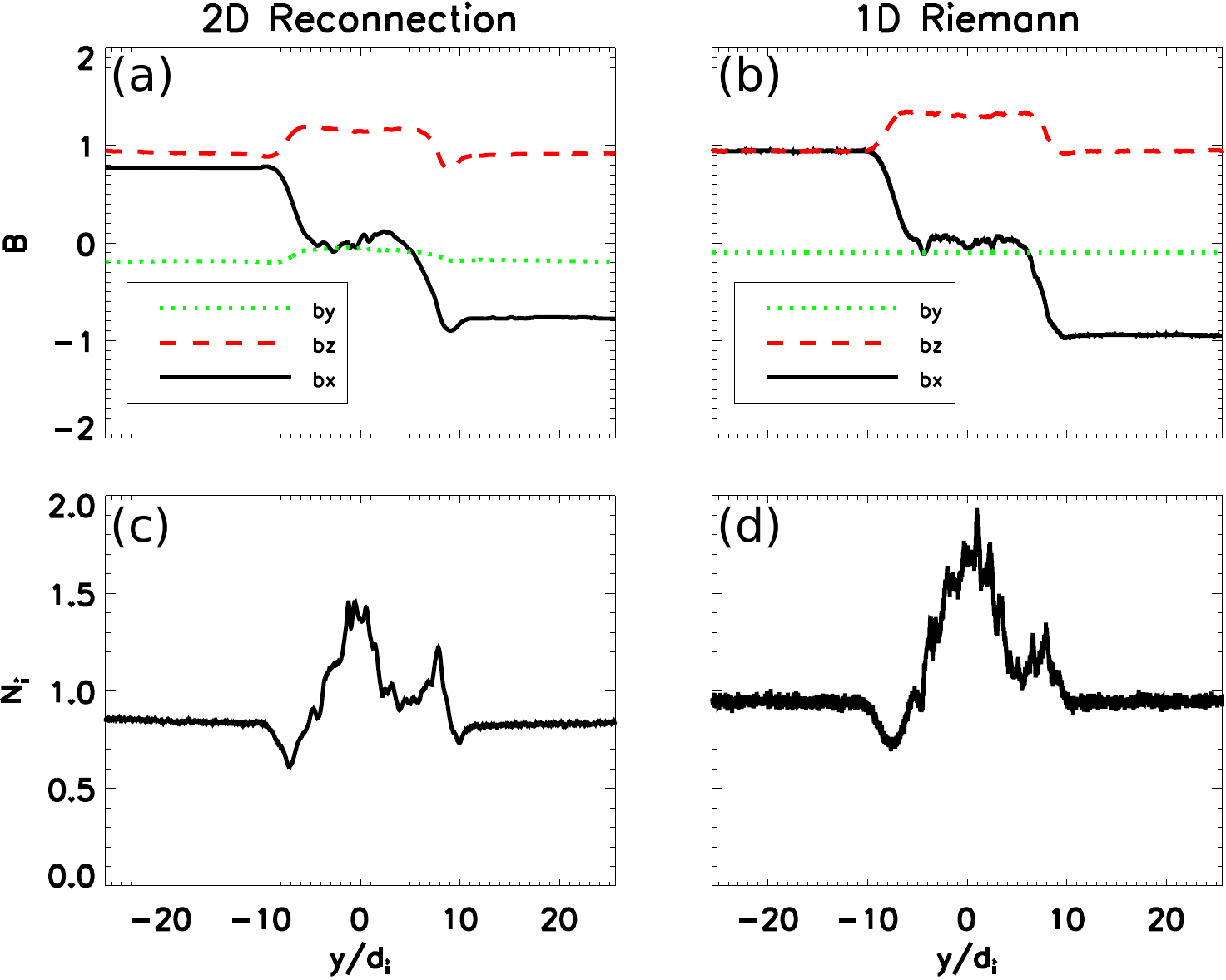}
\caption{\label{fig4} On the left, the profiles of $B_x$, $B_y$, $B_z$ (a) and $N_i$ (c) from the 2D reconnection simulation (Run 1).  On the right, the profiles of $B_x$, $B_y$, $B_z$ (b) and $N_i$ (d) from the 1D Riemann simulation (Run 2).}
\end{figure*}

Here we show that with the same parameters, Riemann simulations
produce comparable results to conventional reconnection
simulations. Some results of a conventional 2D reconnection simulation (Run 1) with a guide field the same as the
reconnecting field are shown in Fig.~\ref{fig1}. All data from Run 1
have been smoothed to reduce the noise. In
Fig.~\ref{fig1}(a), the in-plane magnetic field lines are overplotted
on $J_z$. Well downstream of the x-line, the field lines turn sharply from the $x$ to the $y$ direction, indicating that the
reconnecting field $B_x$ sharply drops to nearly zero. Note, however, that there is a strong guide field $B_z$ so that within the exhaust the magnetic field points dominantly in the $z$ direction.
This feature is characteristic for guide field reconnection but is
absent in antiparallel reconnection where Petschek's switch-off shocks
are suppressed because of pressure anisotropy \cite{Liu2012}. $J_z$
peaks at the exhaust boundaries to support this field change. Between
regions of high current is the exhaust where plasma reaches the
Alfv\'en speed $C_{AX}$, as shown in Fig.~\ref{fig1}(b).  Ions in
the exhaust are heated as shown by the ion parallel temperature
increase shown in Fig.~\ref{fig1}(c). 
In Fig.~\ref{fig2}, we compare this 2D reconnection simulation (Run 1) to
a 1D Riemann simulation (Run 2) with otherwise the same
parameters. The second short dimension $x$ in Run 2 is a dummy
dimension that is included in the simulations but can be averaged to
reduce noise. Fig.~\ref{fig2}(a) shows $T_{e\parallel}$ of the outflow
from the reconnection simulation. The green cut shows the location of
the profiles of parallel electron and ion temperatures shown in
Fig.~\ref{fig2}(b). Fig.~\ref{fig2}(c) shows the corresponding
profiles from the Riemann simulation at the time when the exhaust
width is close to that in Fig.~\ref{fig2}(b). In this paper profiles
along $y$ from Riemann simulations with a short dimension in $x$ have
all been averaged over $x$ to reduce noise. Similarly, we compare
velocity, magnetic field and density profiles in Fig.~\ref{fig3} and
Fig.~\ref{fig4}. The comparable results from both types of simulations
suggest that Riemann simulations are good proxies for the structure of
outflows from conventional reconnection simulations. In the next
section the structure of reconnection outflows will be discussed in
more detail using Riemann simulations.
\section{Results and discussion}
\subsection{Overview} \label{overview}
\begin{figure*}
\includegraphics[width=\linewidth]{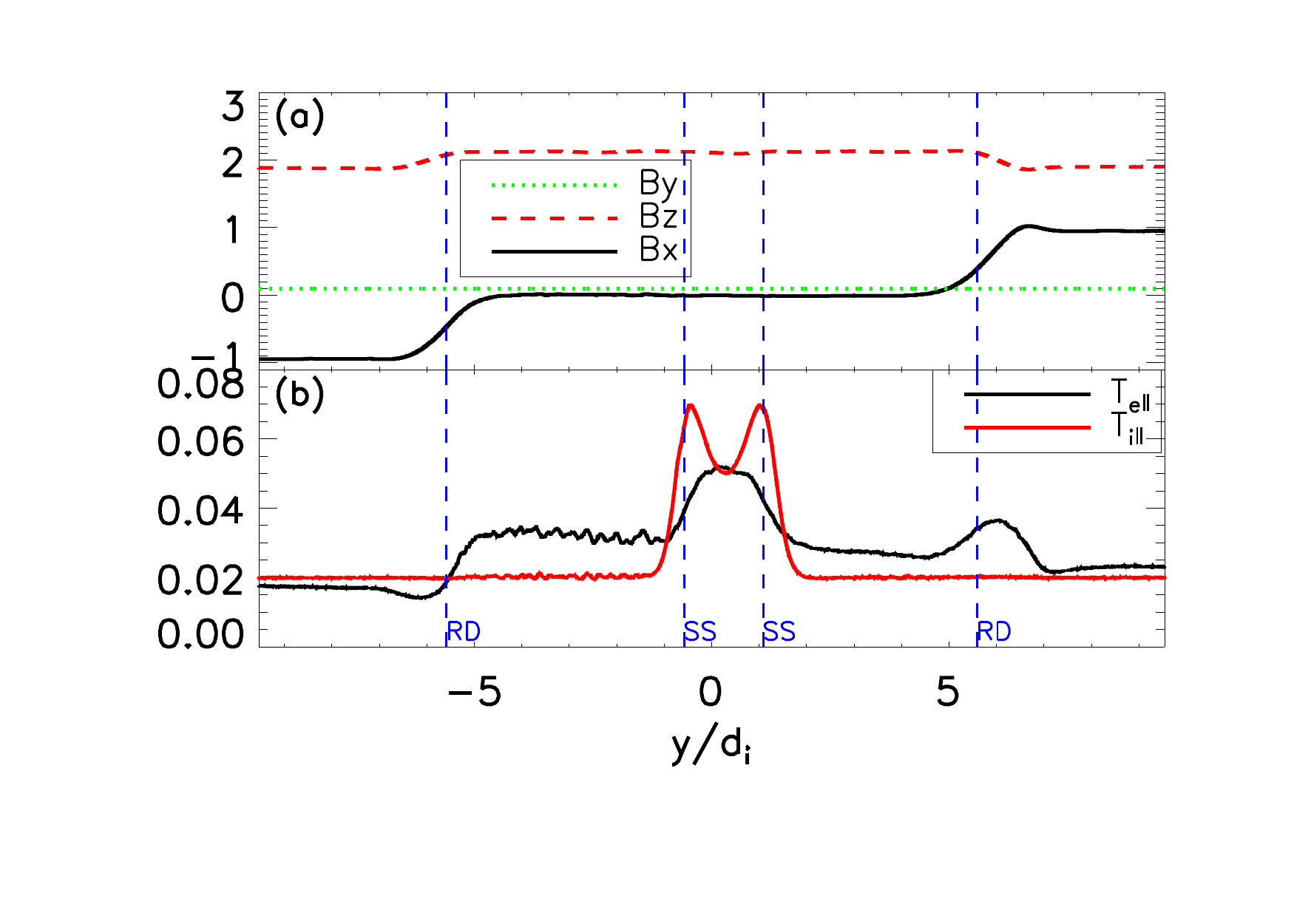}
\caption{\label{fig5} Profiles of $B_x$, $B_y$, $B_z$ (a) and $T_{e\parallel}$ and $T_{i\parallel}$ (b) from the 2D Riemann simulation (Run 3).}
\end{figure*}

\begin{figure*}
\includegraphics[width=\linewidth]{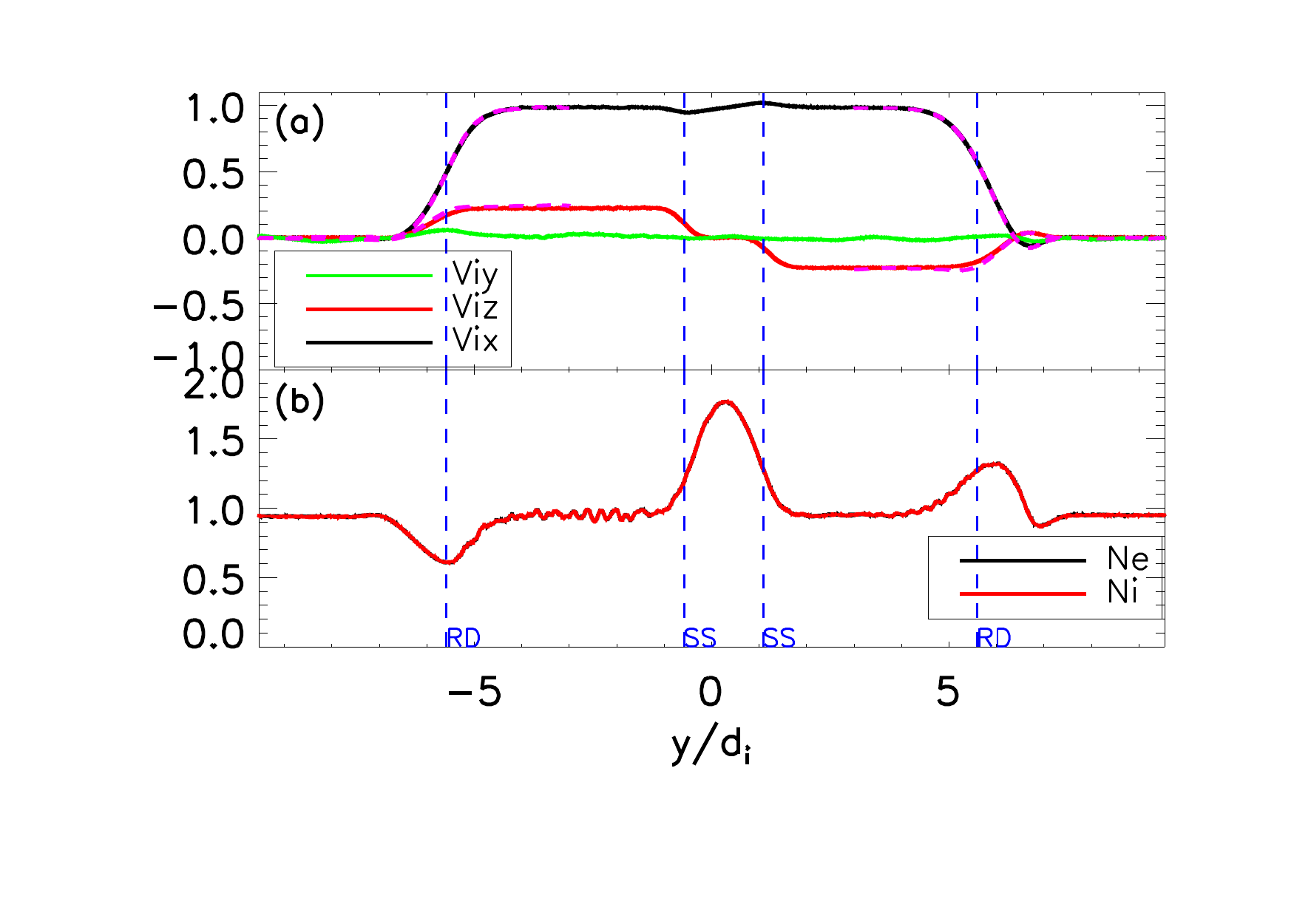}
\caption{\label{fig6} Profiles of $V_{ix}$, $V_{iy}$ and $V_{iz}$ (a) and $N_{e}$, $N_{i}$ (b) from the 2D Riemann simulation (Run 3). The vertical dashed lines indicate the locations of the RDs and SSs. The purple dashed lines in (a) is the MHD model predictions in Equation~(\ref{V_0}) of $V_{ix}$ and $V_{iz}$ for comparison.}
\end{figure*}

In this section, we analyze a 2D Riemann simulation (Run 3), which has
a guide field twice the reconnecting field, in detail to show an
example of typical results. This simulation has a second dimension
along $z$, the dominant magnetic field direction within the exhaust,
that is long enough to capture field-aligned streaming instabilities,
which will be discussed in greater detail later.
After the simulation begins, the ion and electron temperatures in the exhaust increase quickly and reach nearly constant values. Then as the exhaust expands over time, the profiles of temperature and other quantities expand steadily with their shape and magnitude nearly unchanged, resulting in more and more heated particles. Snapshots of the profiles of the magnetic field, the parallel electron and ion temperatures, the bulk flows and electron and ion densities are shown in Figs.~\ref{fig5} and \ref{fig6}.

The expanding exhaust consists of nonlinear structures propagating at
constant speeds away from the initial central current sheet. They are
two rotational discontinuities (RDs), where magnetic fields rotate,
and two slow shocks (SSs), where the fluid velocity decelerates from
above to beneath the slow sonic speed ($\sim 0.2$ upstream of the shock). The structures are
moving away from the midplane at nearly constant speeds. With a
sufficiently large guide field and sufficiently low $\beta$ (in the
case of the guide field equal to the reconnecting field, for example,
$\beta\sim 0.01$), the RDs and SSs are clearly separated. An ideal MHD
Riemann simulation also develops these structures\cite{Lin1993}, but
the detailed properties will differ from those seen here because of
the assumptions in MHD as discussed previously.

In Figs.~\ref{fig5} and \ref{fig6} we present profiles of various
quantities during the exhaust expansion. In Fig.~\ref{fig5}(a)
there is magnetic rotation at each RD (with magnetic field strength
nearly unchanged) with the magnetic fields being nearly uniform throughout the
region between the RDs. In Fig.~\ref{fig5}(b), the strongest ion and
electron parallel temperature increase is between the two SSs but
there is also electron parallel heating between the RD and the SS,
forming two shoulders in the electron parallel temperature
profile. The perpendicular temperature change is negligible due to
magnetic moment conservation and is therefore not shown. In
Fig.~\ref{fig6}(a) $V_{ix}$ increases across the RDs and remains
nearly constant across the entire exhaust, consistent with the MHD
model. $V_{iz}$ increases across each RD with opposite signs on either
side of the exhaust. The resulting counterstreaming flows decreases to
nearly zero across the SSs, again consistent with the MHD model. In
Fig.~\ref{fig6}(b), quasi-neutrality is well satisfied. The density
has a cavity on one RD and a bump on the other one.  The density does
not change much across the RDs, while there is a peak between two
SSs.

We integrate the parallel electric field (smoothed
over one plasma period to reduce fluctuations) to obtain the parallel
electric potential as shown in Fig.~\ref{fig7}(a) and a zoom-in of the
region between the SSs in (b). Note the separate localized variations
of the potential at each RD and SS. The potential gradient drives the
parallel current that produces the magnetic field rotation across the
RDs, maintains zero current elsewhere and enforces
quasi-neutrality in the region between the SSs. These roles will be
discussed in more detail in following subsections. In addition, we
show the parallel phase spaces $y-V_{i\parallel}$,
$y-V_{e\parallel}$ ($V_\parallel=\textbf{V}\cdot\textbf{B}/B$) in
Fig.~\ref{fig7}(c), (e) and the zoom-in of the region between the SSs
in (d), (f). Note that in these figures positive
$V_\parallel$ corresponds to positive $V_y$, and {\it vice versa}.

Before discussing in more detail the structure of the RDs and SSs we
address the role of current-driven instabilities in the low $\beta$
environment considered here. Since the $z$ component of the magnetic
field is the dominant component in the reconnection exhaust between
the two RDs (the x component is nearly zero while $B_y$ remains small
(see Fig.~\ref{fig5}(a)), a long enough $z$ dimension in the
simulations can capture magnetic field aligned streaming
instabilities. The length of the $z$ dimension in our 2D Riemann
simulations is chosen to capture electron-electron, electron-ion or
ion-ion streaming instabilities. The characteristic scale lengths are
$u_b/\omega_{pe}$ for electron-electron and electron-ion instabilites,
and $V_{eth}/\omega_{pe}$ for ion-ion instabilities, where $u_b$ is
the relative velocity between two beams and $V_{eth}$ is the electron
thermal speed\cite{Krall1973, FUJITA1977}. In Fig.~\ref{fig8}, we show
the parallel electric field $E_\parallel=\textbf{E}\cdot\textbf{B}/B$
in the $y-z$ plane of the Run 3 simulation listed in Table
\ref{table1}. There is evidence for instability at each RD (especially
at the left RD), but there is no instability around or downstream of
the SSs. We focus on the left RD, which exhibits a stronger
instability. The turbulence is produced by the Buneman instability
driven by the electron beam supporting the current at the RD. Since
the width of the RD in the simulation has a $d_i$ scale, from Ampere's
law, the beam speed is on the order of $B_{x,u}c/4\pi ned_i=C_{Ax,u}$, the
Alfv\'en speed. So the instability is expected to become weaker with higher
mass ratio due to the higher electron thermal speed relative to Alfv\'en
speed. In Fig.~\ref{fig9}, we compare the instability in the current
run (Run 3) with mass ratio 400 to that in Run 4 with mass ratio 100. We see
that the instability is significantly weaker in the higher mass ratio
simulations. Further, from the electron phase space in
Fig.~\ref{fig7}, we see that the instability does not significantly
limit the electron beam at the left RD. Thus, the instabilities do not
play a significant role either in the region around the SSs or the
RDs. The driver for the instabilities and their impact on the exhaust
profile will be discussed in greater detail in a follow-up paper.

\subsection{Rotational discontinuity (RD)} \label{RD}
Across an RD ions undergo a jump in velocity that can be calculated from the MHD model \cite{Lin1993}.   
In the limit of low upstream $\beta$,
\begin{equation}\label{V_0}
\begin{split}
V_{x,d}=s\frac{B_{x,d}-B_{x,u}}{\sqrt{4 \pi n_0 m_i}} \\
V_{z,d}=s\frac{B_{z,d}-B_{z,u}}{\sqrt{4 \pi n_0 m_i}}
\end{split}
\end{equation}
where the subscripts $u$, $d$ designate upstream and downstream of the RD and $s=sgn(V_{yu} B_{yu})$, all evaluated in the frame of the RD.
Equation~(\ref{V_0}) agrees well with simulations carried out with sufficiently low upstream $\beta$. It can not only predict the total jump across the RD but also the continuous transition across the RD, if the downstream magnetic field is treated as a continuous function. As shown in Fig.~\ref{fig6}(a), the purple dashed lines are consistent with the velocity profiles. 
The ion velocities in x and z directions are driven by magnetic tension in $x$ and $z$.
Equation~(\ref{V_0}) indicates that $V_z$ downstream of the RDs has opposite signs on either side of the midplane as seen in Fig.~\ref{fig6}(a). This leads to two ion beams traveling towards the center along the magnetic field with $V_0 \approx |V_{z,d}|$, since $B_z$ is the dominant magnetic field component between the two RDs. These two beams counterstream and give rise to the two SSs. 
Note that in Fig.~\ref{fig6}(a), $V_x$ is symmetric because $B_x$ and $s$ are anti-symmetric. $V_z$ is anti-symmetric because $B_z$ is symmetric while $s$ is anti-symmetric.

\begin{figure*}
\includegraphics[width=\linewidth]{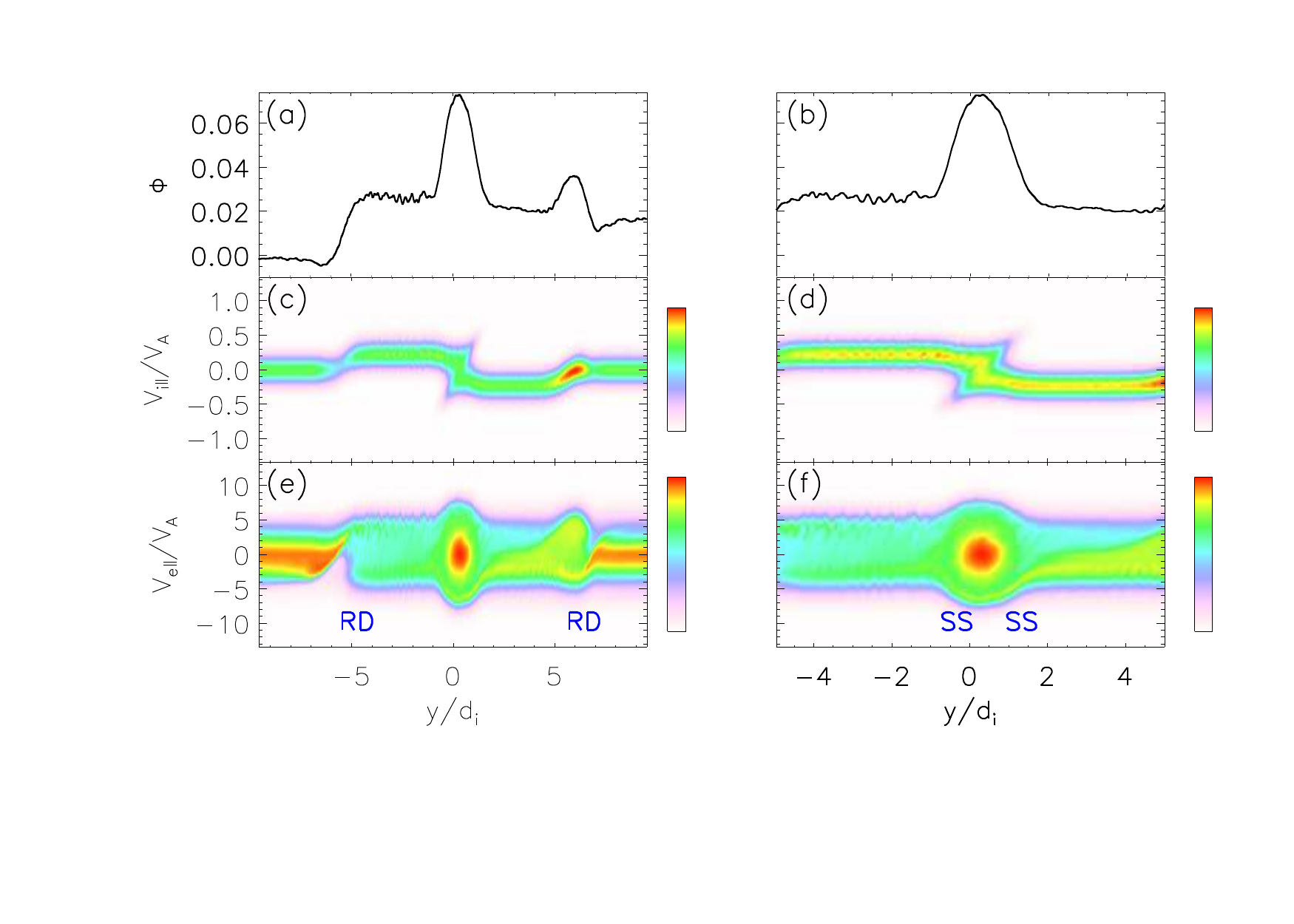}
\caption{\label{fig7} On the left, from top to bottom, the profile of parallel electric potential $\phi$ (a), ion phase space $y-V_{i\parallel}$ (c) and electron phase space $y-V_{e\parallel}$ (e) across the whole exhaust. On the right, the same quantities (b) (d) (f) zooming in on the region between the SSs.}
\end{figure*}

\begin{figure*}
\includegraphics[width=\linewidth]{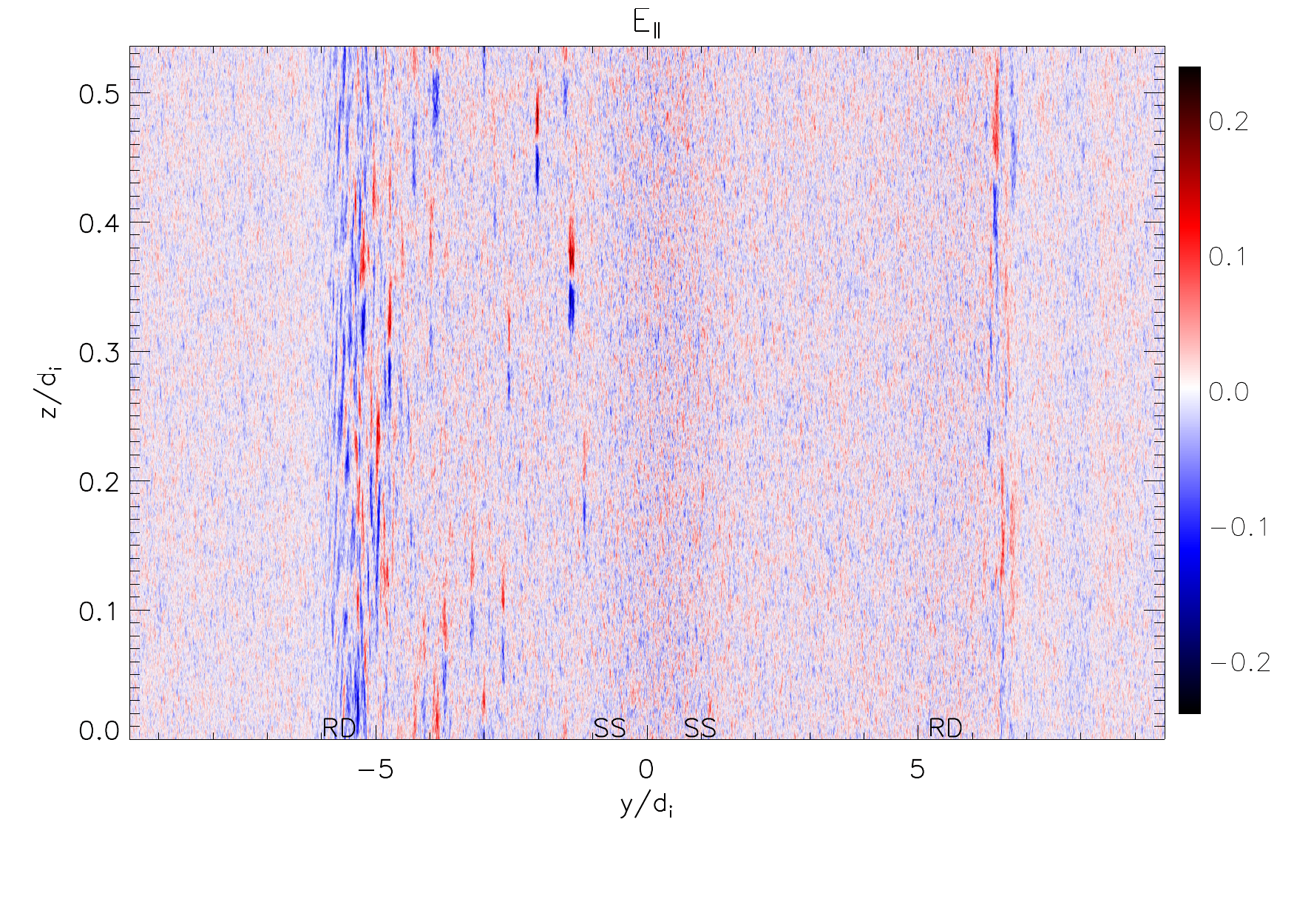}
\caption{\label{fig8} Parallel electric field $E_\parallel$ from the 2D Riemann simulation (Run 3). Note the different axis scales.}
\end{figure*}

\begin{figure*}
\includegraphics[width=\linewidth]{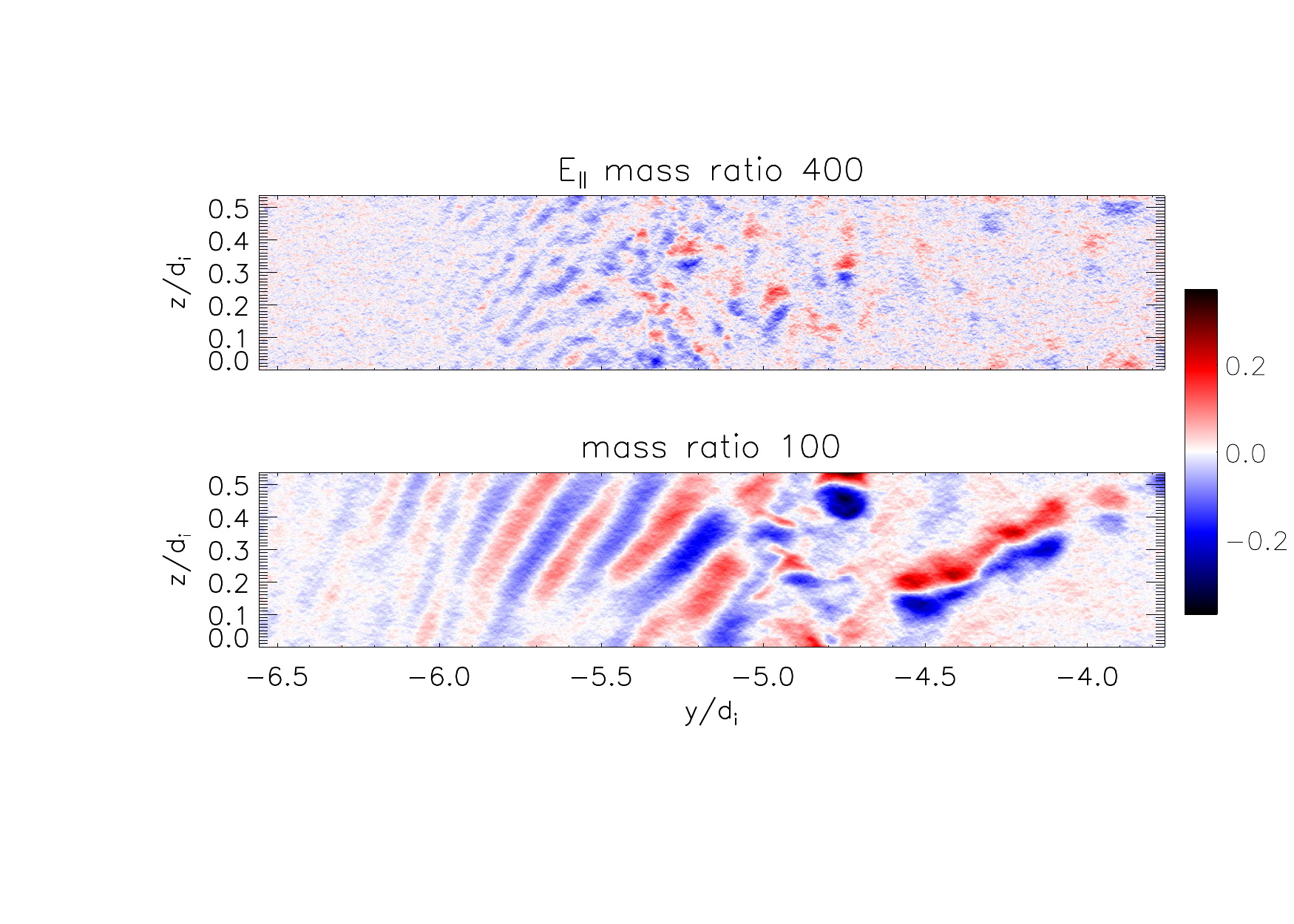}
\caption{\label{fig9} Parallel electric field $E_\parallel$ of the left RD region from two 2D Riemann simulations Run 3 (top) and Run 4 (bottom) at the same time. The two figures have the same spatial and color scales.}
\end{figure*}

\begin{figure*}
\includegraphics[width=\linewidth]{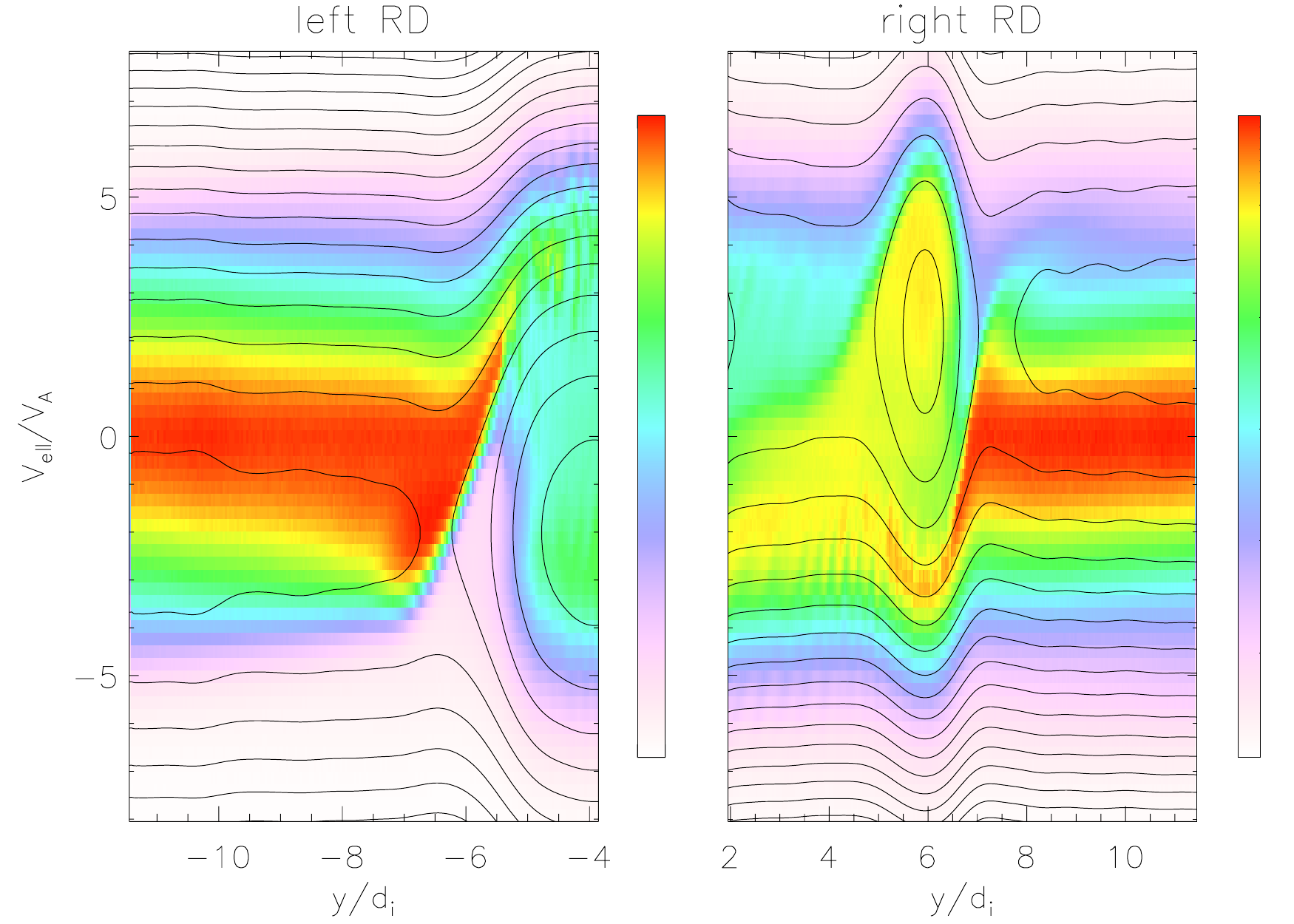}
\caption{\label{fig10}  The phase space $y-V_{e\parallel}$ of the regions around the RDs on the left and right of the exhaust with the contours of parallel mechanical energy evaluated using the potential at this time in the frame of the RD potential ramp at the outer edge to show the approximate phase space stream lines of electrons under the potential.}
\end{figure*}

While the ion motion across the RDs is controlled by magnetic tension, the electrons are controlled by the localized parallel potentials at the RDs. As a result of these potentials, the electron distributions carry a localized parallel current at the RDs to support the magnetic field rotation while maintaining zero current elsewhere. This leads to partial electron confinement within the exhaust. We demonstrate this in Fig.~\ref{fig10}. We show the phase space $y-V_{e\parallel}$ of the RD regions on the left and right of the exhaust and overplot the contours of parallel mechanical energy evaluated in the frame of the RD at the outer edge of exhaust. The mechanical energy is obtained by calculating $\frac{1}{2}m_e(V_{e\parallel}-V_{ramp\parallel})^2-e\phi$, where $\phi$ is calculated as in Fig.~\ref{fig7}(a) and $V_{ramp\parallel}$ is the effective speed of the potential ramp along the magnetic field seen by the electrons. So $V_{ramp\parallel}=V_{ramp,y}B/B_y$ where $V_{ramp,y}$ is the ramp speed relative to the ${\bf E}\times {\bf B}$ drift in y direction at the ramp. We measure $V_{ramp\parallel}$ to be -2.0 for the left RD and 2.2 for the right RD. 
In Fig.~\ref{fig10} we see that the electrons mostly follow the
stream lines, suggesting the potential is controlling the electron
motion. In this phase space electrons with positive (negative)
$V_{e\parallel}$ at the left (right) RD are streaming toward the
midplane of the exhaust. As electrons enter the left RD from upstream
(positive $V_{e\parallel}$) a small potential dip reflects some of the
low-velocity electrons. The dominant potential (see Fig.~\ref{fig7}(a))
then accelerates the incoming electrons across the RD, driving a
localized current at the RD that supports the magnetic field
jump. Most of the electrons moving toward the left RD from within the
exhaust are reflected by the potential at the RD and therefore are
effectively confined within the exhaust.
At the right RD, the
downstream outgoing electrons are first accelerated towards upstream
and then decelerated to produce the localized current that supports the magnetic field jump at the RD (see the potential in
Fig.~\ref{fig7}(a)). Some of these electrons leak out of the exhaust, while some are reflected back towards the midplane. Similarly, some of the incoming electrons are accelerated into the exhaust and then decelerated. Other incoming electrons (negative
$V_{e\parallel}$) are reflected back upstream by a small potential dip. 
Comparing both RDs, more of the downstream electrons leak across the right RD to the
upstream than across the left RD.  Thus, a higher
fraction of electrons are confined by the RD where the electric field
driving the current at the RD acts as a confining electric field. The
electron confinement at either side helps to maintain zero current
upstream.
In the regions between the RD and the SS on either side, as in
Fig.~\ref{fig7}(e), there are electrons from the RD and electrons that
have escaped from the region between the two SSs. The multiple
electron populations between the RD and the SS contribute to a
somewhat higher electron temperature than upstream, which is seen at
the shoulders in Fig.~\ref{fig5}(b). There is no counterpart to these
shoulders in the MHD model.

Electron confinement at the edge of the exhaust was also observed in
simulations reported by Egedal et al.\cite{Egedal2015,Egedal2012}. Their reconnection simulations were in
the low-$\beta$, anti-parallel regime. They found almost complete
electron confinement on both sides of the exhaust in the region just
downstream of the x-line. This was a consequence of a large potential
which was driven by the magnetic expansion and ion demagnetization
near the x-line. This mechanism, however, is not active far downstream
of the x-line where ions are magnetized. Further, in guide field reconnection
magnetic expansion is suppressed. As a consequence, we do not see
such a large confinement potential develop, especially at the right RD.
\subsection{Slow Shock (SS)}
In the region between the SSs, the dynamics of both ions and electrons
are controlled by the parallel potential. As shown in
Fig.~\ref{fig6}(b), upstream of the shock both ions and electrons have
the same density, which is close to the ambient density $n_0$ upstream
of the RDs. In Fig.~\ref{fig7}(d), ions moving from upstream to
downstream across the SSs are decelerated with a small fraction
reflected. Some faster ions reach the SS on the other side of the
exhaust and are accelerated into the region upstream of the SS.
The counterstreaming ion beams around and between the SSs
increase the effective ion temperature although the distributions
retain beam-like features. 

In contrast with the ions, the electrons are accelerated downstream
across the SSs (Fig.~\ref{fig7}(f)). Since the SSs are moving outward,
some lower energy electrons are trapped by the retreating potentials and
lose energy over time due to conservation of the second adiabatic
invariant as the region between the SSs expands. Other higher energy
electrons have high enough energy to go through the potential to
escape from the region between the two SSs. The trapped electrons
result in the higher electron temperature downstream of the SSs. Since
it is the ion beams that are the energy source of the SSs, the
electron heating represents the conversion of ion bulk flow energy to
the electrons. Note that in the electron phase space shown in
Fig.~\ref{fig7}(f), there is a localized peak near (y=0,
$V_{e\parallel}$=0) on the top of the rest of distributions with the
maximum phase space density close to the initial distribution
maximum. This is a trapped population left over from the initial
formation of the RD and SS. These trapped electrons lose energy as the
exhaust expands and become energetically unimportant at late time.

The two SSs are formed by the counterstreaming ion beams produced at
the RDs (see Fig.~\ref{fig6}(a)). In the frame of the exhaust
downstream of the RDs the beams propagate along the nearly constant
magnetic field (see Fig.~\ref{fig5}(a)) so the resulting SSs are
electrostatic shocks. The charge imbalance driven by the beams
produces the jump in the parallel potential across the SSs. If there
were no potential, the counterstreaming ion beams would produce an ion
density of $2n_0$ in the central region. In contrast, due to high
electron thermal velocity, only half of the electrons from either side
would reach the region with counterstreaming ions. The remaining half
of the electrons would never reach the region of counterstreaming
ions. Thus, in the absence of the potential, the central electron
density would be only $n_0$. The charge imbalance between ions and
electrons drives the potential, which modifies the distribution
functions of both species and restores quasineutrality. In the low
initial $\beta$ limit of the anisotropic MHD model as is discussed in
the Appendix, the speed of the SS along the magnetic field is close to 
$V_0$, just like a gas dynamic shock. This speed matches the results of 
simulations with sufficiently low $\beta$. 
If the inflowing
distributions of ions and electrons into the region between the SSs
were known, one could use Liouville's theorem to kinetically express
the ion and electron distributions at the center downstream of the two
SSs as a function of the  potential jump across the shock, which would yield their densities. Using quasineutrality one could then equate the
densities of ions and electrons to solve for $\phi$ and use it to
determine the central distribution functions, densities and
temperatures. Thus, it is quasineutrality that controls the magnitude
of the potential and the dynamics of ions and electrons. However, the
major difficulty with this method is that the inflowing electron
distributions into the SS from the RD are nontrivial (as discussed in
the previous subsection). We will further discuss the quasineutrality
requirement in the low-$\beta$ regime in the next section.

\section{Scaling of heating and energy partition in the low-$\beta$ regime}
\subsection{Justification of 1D Riemann simulations}
\begin{figure*}
\includegraphics[width=\linewidth]{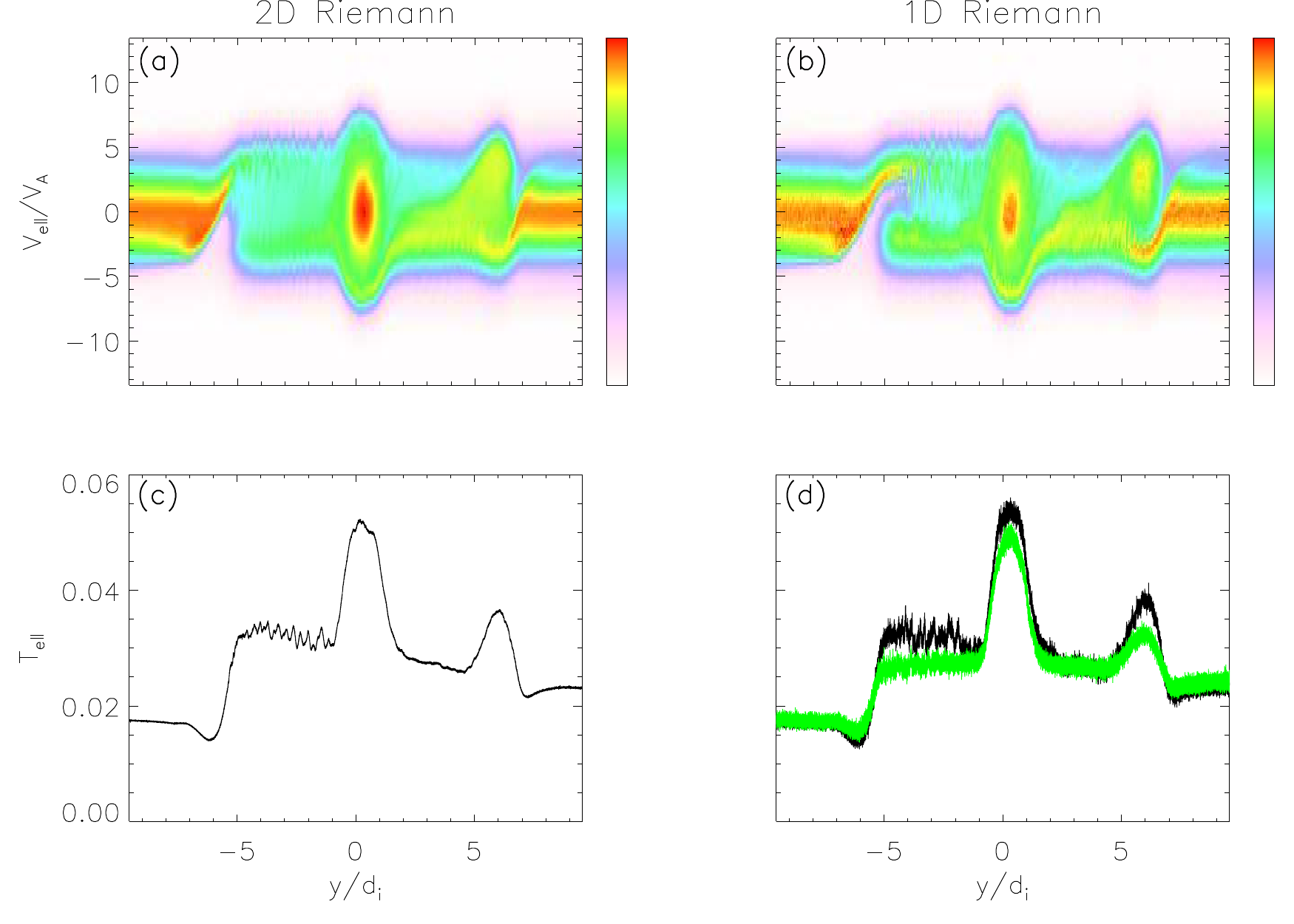}
\caption{\label{fig11} Comparing a 2D Riemann simulation and a 1D Riemann simulation. On the left from the 2D Riemann simulation (Run 3), the electron phase space $y-V_{e\parallel}$ (a) and the profile of $T_{e\parallel}$ (c). On the right the corresponding quantities (b) (d) from the 1D Riemann simulation (Run 5). The green line in (d) is the $T_{e\parallel}$ profile from Run 13 with mass ratio 1600.}
\end{figure*}

\begin{figure*}
\includegraphics[width=\linewidth]{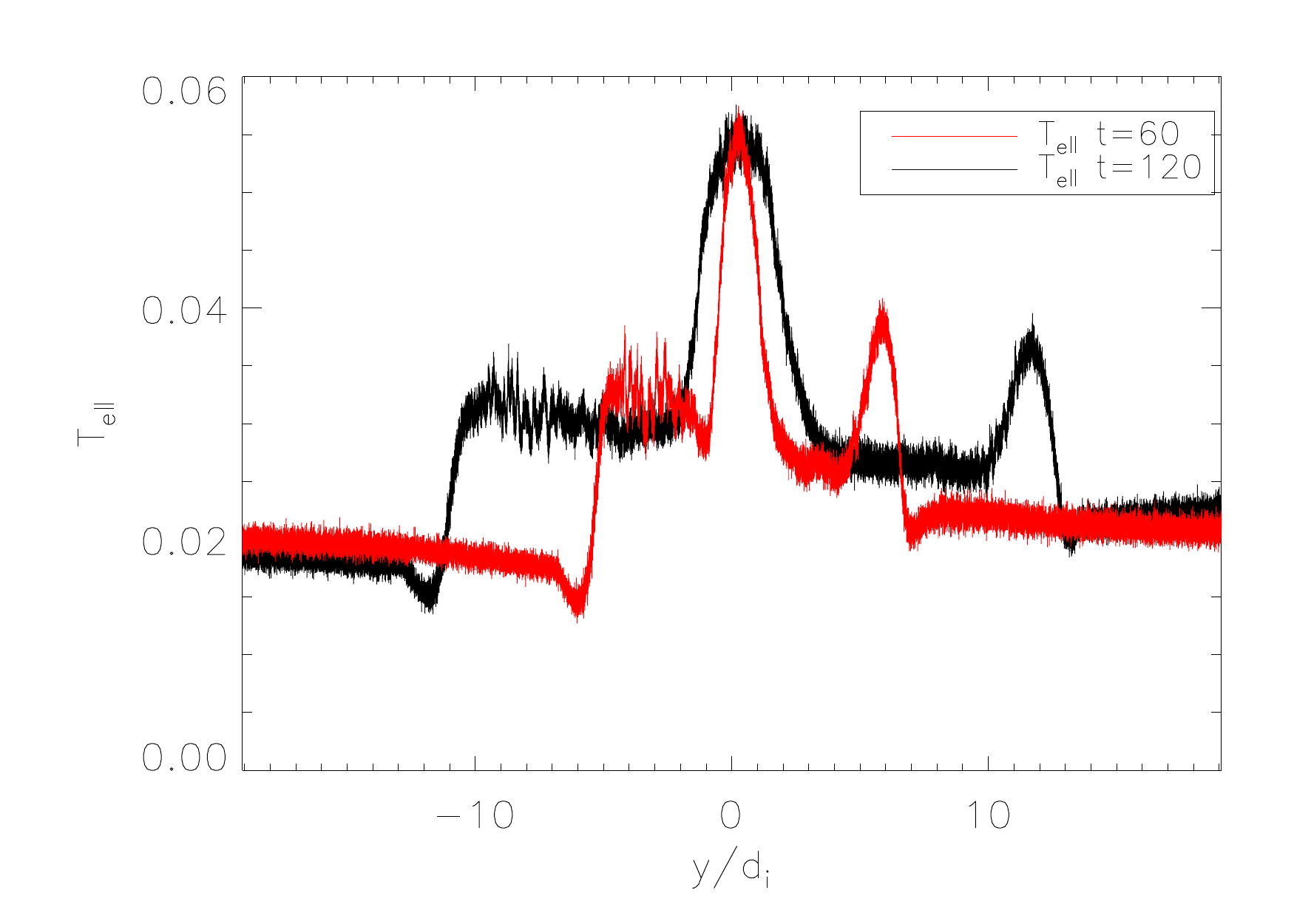}
\caption{\label{fig12} The parallel electron temperature profiles of Run 14 at t=60 and t=120.}
\end{figure*}

To explore the scaling of ion and electron heating in the low-$\beta$
regime we perform a series of 1D Riemann simulations.  By ignoring the
$z$ direction, we eliminate the possible development of streaming
instabilities such as those seen in Fig.~\ref{fig8}. However, these
instabilities have little effect on the system's development. To
demonstrate this we show in Fig.~\ref{fig11} a comparison between a 2D
Riemann simulation in the y-z plane (Run 3) and a 1D Riemann (Run 5)
simulation based on the same parameters. Panels a and b show the phase
spaces and panels c and d (black line) show the $T_{e\parallel}$
profiles.  The similarity of the panels suggests that the eliminated
instabilities that did develop in the 2D simulation are too weak to have a
significant impact on the results. Also we show in panel d (green line) the
$T_{e\parallel}$ profile from Run 13 with a mass ratio 1600 and
otherwise the same physical parameters as Run 5 to demonstrate that
the results are not sensitive to the mass ratio as long as it is high
enough. In addition, we perform a 1D Riemann simulation (Run 14) doubling the domain size in y of Run 5, so that we can double the simulation time from 60 to 120. We show the electron parallel temperature profiles at t=60 and t=120 in Fig.~\ref{fig12}. We demonstrate that the structures and heating remains the same as the exhaust further expands over time. Hence, in the next section we will use 1D Riemann simulations to scan the low-$\beta$ regime.
\subsection{Scaling of electron and ion heating with the released magnetic energy}
\begin{figure*}
\includegraphics[width=\linewidth]{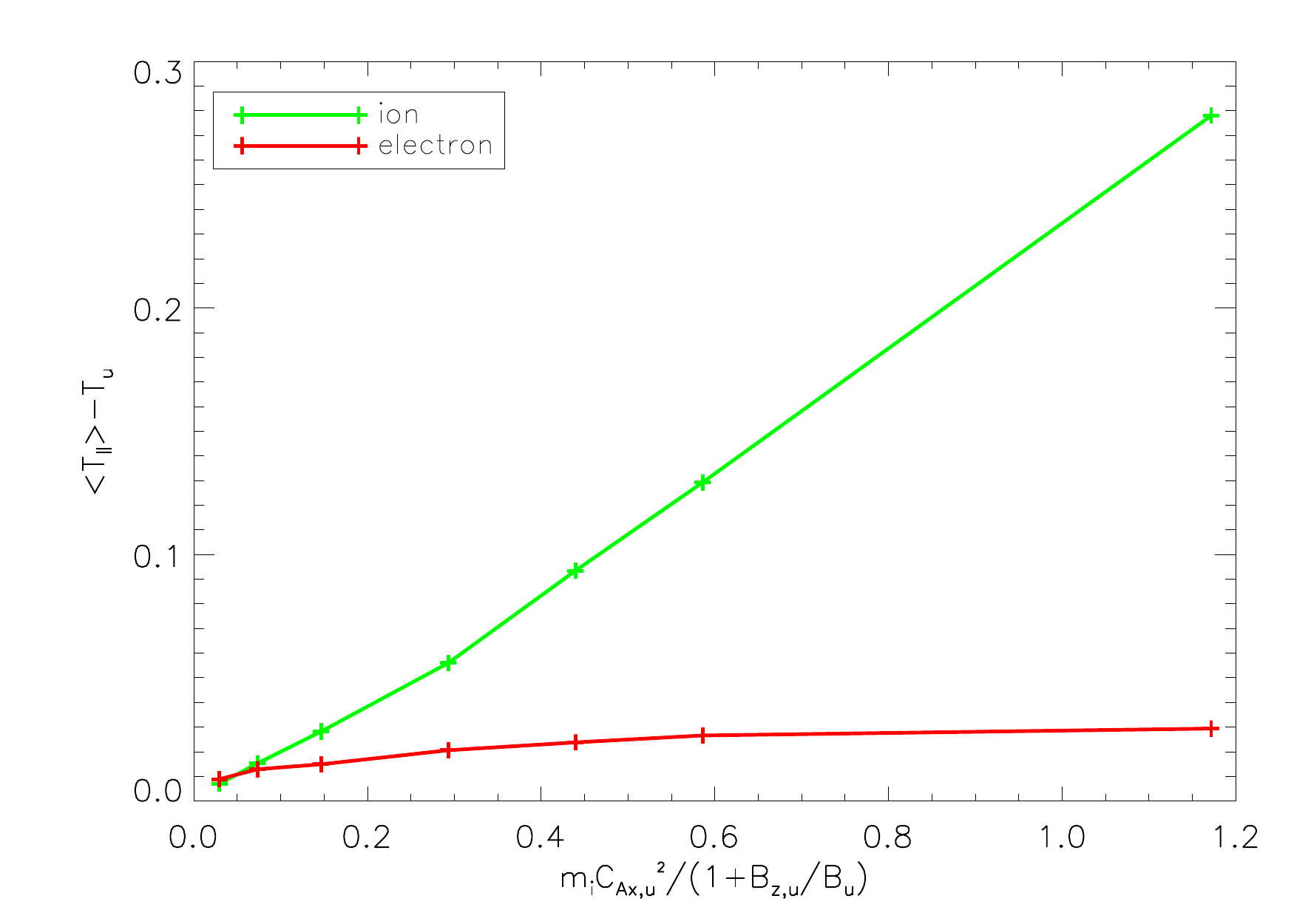}
\caption{\label{fig13} The scaling of parallel heating of ions and electrons as a function of available magnetic energy per particle using data from Runs 5-10}
\end{figure*}

Here we present a series of simulations (Runs 5-11) in which the only difference in the initial profiles of the physical quantities are the magnitudes of the upstream  magnetic fields. For these runs the electron $\beta$ varies between 0.1 and 0.0025. 
For the lower $\beta$, the higher mass ratio is needed to ensure that the electron thermal speed exceeds the characteristic ion flow, RD and SS effective speeds along the field, etc. The requirements on the mass ratio will be discussed more in the next subsection. 
In Fig.~\ref{fig13}, we plot the variation of the ion and electron temperature increase averaged over the region between the two SSs. The horizontal axis is the available magnetic energy per particle in the low-$\beta$ limit $m_iC_{Ax,u}^2/(1+B_{z,u}/B_{u})$ derived from anisotropic MHD (see the Appendix).

We see in Fig.~\ref{fig13} that the ion heating is proportional to the available magnetic energy per particle in the low $\beta$ limit as expected, while the electron heating reaches a plateau in the low-$\beta$ limit. In contrast, previous observational and computational reconnection scaling studies suggest that the electron heating should exhibit a linear scaling\cite{Phan2013,Shay2014}. However, these previous studies only focused on the $\beta$ of order unity regime and therefore did not reach low enough $\beta$ to see the saturation of the electron heating. We find one simulation (number 302) in Shay's paper\cite{Shay2014} with both initial ion to electron temperature ratio and guide field to reconnecting field ratio equal to one that can be compared with one of our simulations. We confirm that our highest $\beta$ run (Run 6) produces comparable electron heating to Shay's simulation if we renormalize our run's available magnetic energy per particle to be the same as Shay's run and we calculate the heating averaged over the whole exhaust as Shay did. Therefore, the Riemann simulation results here are consistent with the previous results at higher $\beta$. 
The physical reason for the saturation of electron heating with available magnetic energy is discussed in the next subsection. The consequence is that the ion heating dominates over electron heating in the limit of low upstream $\beta$.

The SS potential can be evaluated by integrating the electron parallel momentum equation across the SS neglecting the inertia term\cite{Haggerty2015}, 
\begin{equation}
\label{phi}
e\Delta\phi= \Delta T_ {e\parallel} +
  \int dsT_ {e\parallel} \nabla _ {\parallel}\ln (n) + \\ \int ds(T_ {e\parallel}- T_ {e\bot}) \nabla _ {\parallel}\ln (B),
\end{equation}
with $ds$ the distance along the local magnetic field. The third term
on the right can be neglected because the magnetic field is nearly
constant across the SS. The potential therefore scales like the
electron temperature. Since this is small in the low $\beta$ limit,
the potential is insufficient to significantly alter the velocity of
the ions as they cross the SS. The consequence is that ion reflection
by the shock potential does not take place, which eliminates the
reflected ion beams upstream of the slow shock that play such an
important role in high Mach number parallel shocks.
Downstream of the SS the ions remain as distinct counterstreaming
beams with essentially no mixing. Although the counterstreaming ion
beams have significant free energy, the ion-ion two stream instability
along the field lines is stable since the electron temperature is
low with the consequence that the ion beam speed is higher than the sound speed
$\sqrt{T_{e\parallel}/m_i}$\cite{FUJITA1977}. As a crosscheck we carried out a 2D
test simulation with uniform magnetic fields and parallel
counterstreaming ion beams with speed higher than sound speed. We did
not observe any instabilities develop to release the energy
associated with the counterstreaming ion beams. This result is consistent
with Fujita et al.\cite{FUJITA1977}.
\subsection{The saturation of electron heating at low $\beta$}
\begin{figure*}
\includegraphics[width=\linewidth]{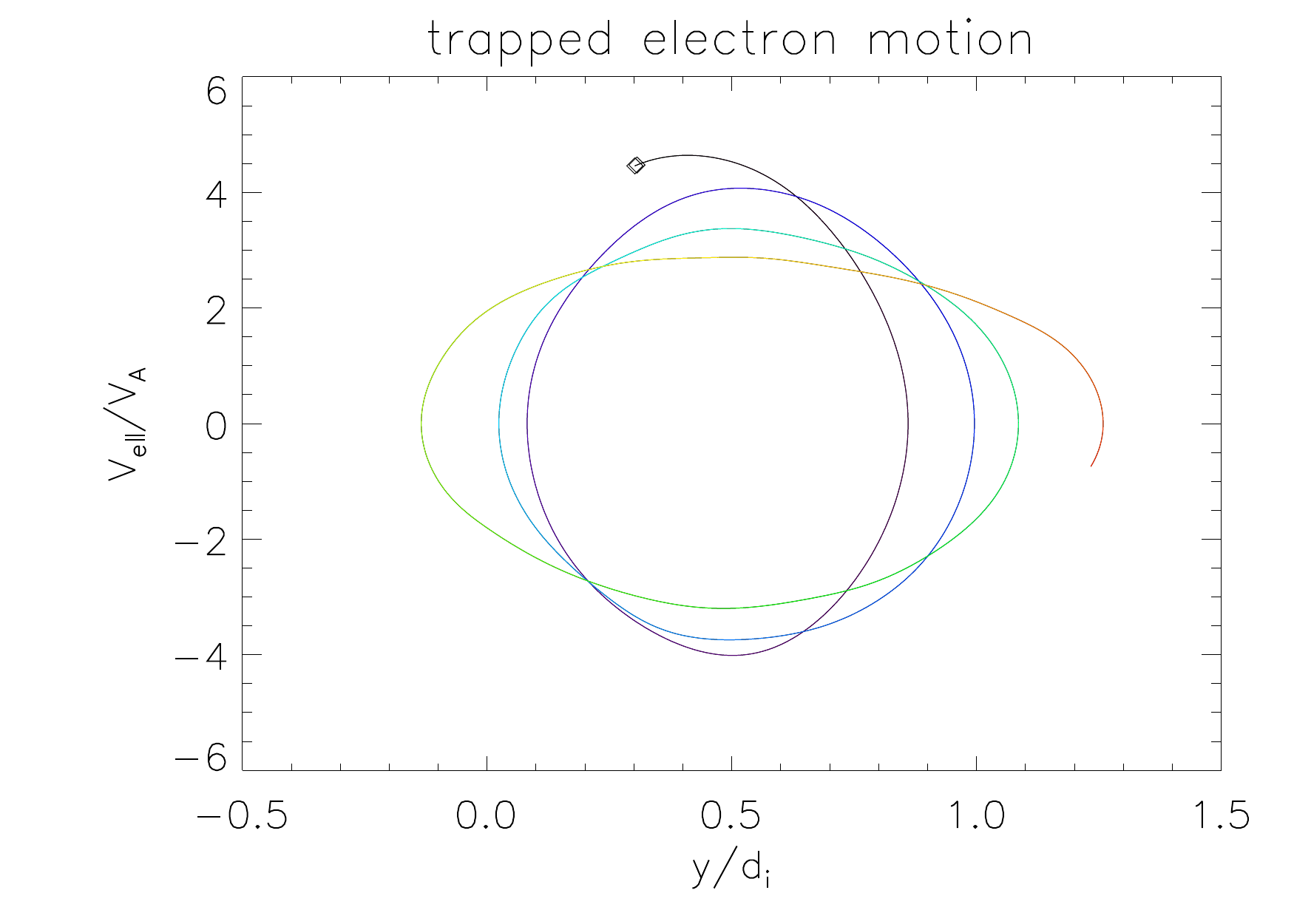}
\caption{\label{fig14} A test particle trajectory in $y-V_{e\parallel}$ phase space using the smoothed magnetic fields and parallel electric potential from Run 11.  The trajectory begins at the diamond and changes color from black to red during the particle motion.}
\end{figure*}

Here we discuss the physics behind the saturation of electron heating
in the low $\beta$ regime. The electron parallel temperature increase
across the RD in our simulations is small because the electron thermal
speeds are much higher than the streaming velocities at the RDs
required to form the current needed to switch-off $B_x$. As the
electrons downstream of the RD cross the SS, the electrons gain energy
because of the high potential between the SSs.  Electrons below a
critical speed $V_{trap}$ in the lab frame will get trapped between
the SSs, while those above it will free stream across both SSs to the
other side of the exhaust.  We evaluate $V_{trap}$ in the
following. We first point out that an electron with this critical
velocity upstream of the first SS will, after passing through both
SSs, reach zero velocity in the frame of the second. We trace an electron with 
 zero velocity just outside of the second SS backwards in time. 
Before crossing the second SS this electron has a velocity
$V_\phi=\sqrt{2\phi/m_e}$ in the frame of the SS. Switching to the frame of the first potential, its parallel velocity is $V_\phi+2V_s$ where $V_s$ is the effective speed of
the SS along the magnetic field in the lab frame. In this frame
before crossing the first potential, the speed is
$\sqrt{(V_\phi+2V_s)^2-V_\phi^2}=2\sqrt{V_sV_\phi+V_s^2}$. Now
changing back to the lab frame, we obtain the critical velocity
$V_{trap}=2\sqrt{V_sV_\phi+V_s^2}-V_s$.  The trapped
electrons then undergo adiabatic deceleration in the expanding trap. We
demonstrate this in Fig.~\ref{fig14} using a test particle
trajectory in the phase space $y-V_{e\parallel}$. Here we have applied
the time dependent background profiles of magnetic fields and smoothed
parallel electric potentials from Run 11. The potential profile is
obtained from equation~(\ref{phi}), which is close to that from
directly integrating parallel electric fields as in
Fig.~\ref{fig7}(a). The particle starts at the diamond point and moves
from black to red color over time, decelerating towards zero velocity.

From charge neutrality the flux of ions and electrons that remains between the SSs must be equal.  In the low $\beta$ limit, the
upstream ion is an incoming beam with speed $2V_s$, so the incoming
ion flux is $2n_0V_s$. All of these ions remain between the two
SSs. The trapped electrons make up the dominant
component of the downstream electrons since the untrapped electrons transit out of the region between the SSs very quickly.  Thus, the incoming flux of electrons that will be trapped must match the total incoming flux of ions. We take the upstream thermal speed
$V_{eth,u} \gg V_s$ (so $\beta$ can not be too low) and $V_\phi\gg V_s$ so
$V_{trap}$ simplifies to $2\sqrt{V_sV_\phi}$. The
upstream electrons with velocities between v=0 and $2\sqrt{V_sV_\phi}$ will
be trapped. Taking $V_{eth,u} \gg V_{trap}$, the fraction of trapped electrons is 
$V_{trap}/V_{et,u}$. The
electron flux is then given by $n_0V_{trap}^2/2V_{eth,u}$. Equating the incoming fluxes of the two species, we have
\begin{equation}
n_0\frac{V_{trap}^2}{2V_{eth,u}}\sim 2n_0V_s
\end{equation}
to obtain $V_\phi\sim V_{eth,u}$ or $e\phi\sim T_{e,u}$.  Therefore,
$T_{e,d}\sim e\phi\sim T_{e,u}$.  Thus, the electron heating can not
be very strong even with large available magnetic energy per particle
and the electron heating reaches a plateau as shown in
Fig.~\ref{fig13}. Physically, this is because the electron heating is
limited by the amplitude of the potential across the SSs. A very large
potential does not develop because it would trap too many electrons
compared with the modest increase in ion density (a factor of two) and
so charge neutrality would be violated.

The conditions used above, $V_{eth,u}\sim V_\phi\gg V_s$, are satisfied in our simulations as long as the ion-to-electron mass ratio is sufficiently large.
\subsection{Partitioning of the ion energy gain}

\begin{figure*}
\includegraphics[width=\linewidth]{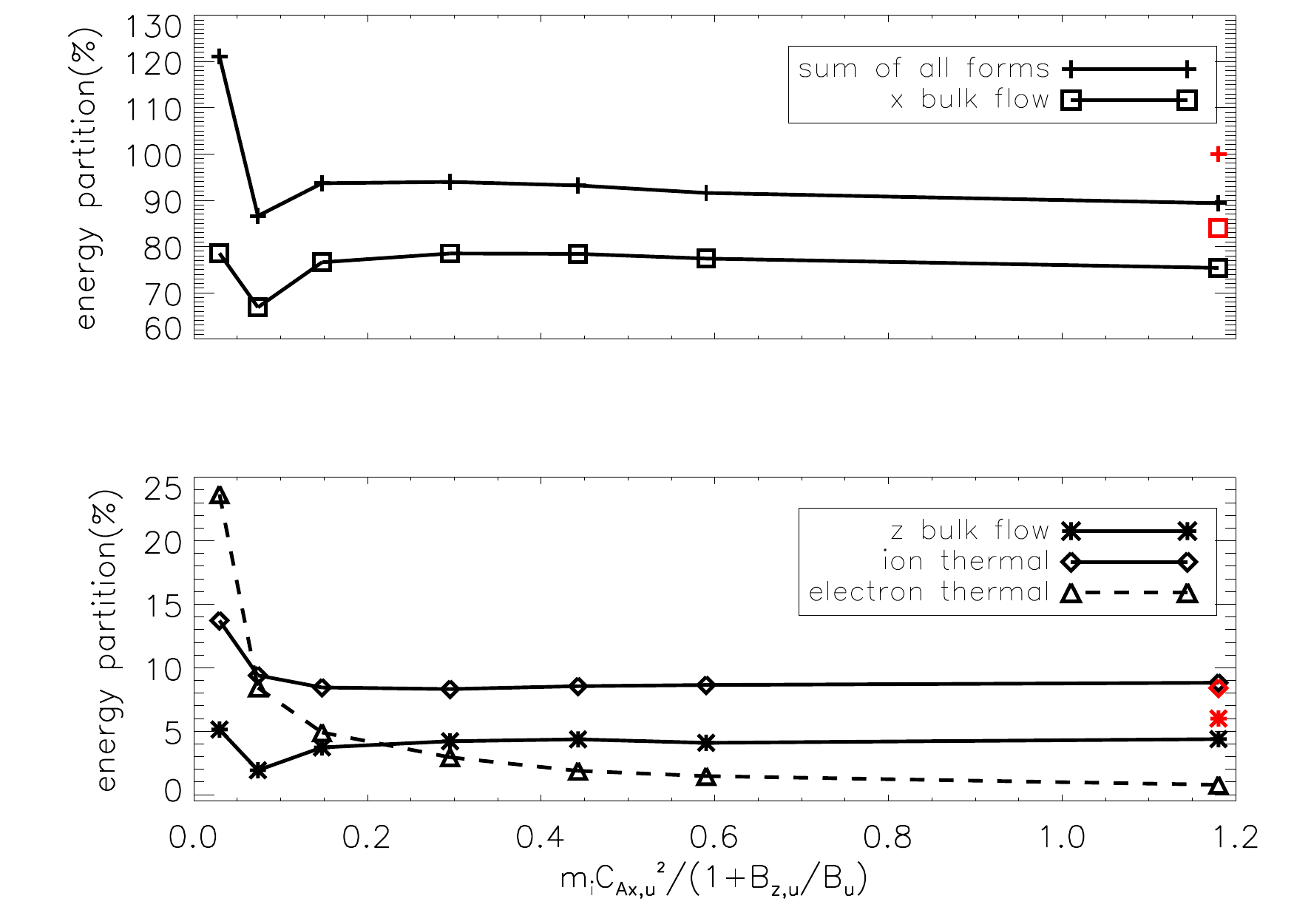}
\caption{\label{fig15} Energy partition into different forms of kinetic energy as a function of available magnetic energy per particle. The corresponding predicted partition by anisotropic MHD in the low initial $\beta$ limit is plotted in red}
\end{figure*}

Magnetic energy flows into the exhaust and is converted into different
forms of energy. As expected from the dominance of the ion temperature in
Fig.~\ref{fig13}, in the low-$\beta$ limit the ion thermal energy dominates the electron thermal energy. In
this limit, the electron thermal energy upstream and downstream can be
neglected. The ion energy gain across the exhaust can be calculated
using the anisotropic MHD solution in the Appendix. The available
magnetic energy per particle was calculated previously to be
$m_iC_{Ax,u}^2/(1+B_{z,u}/B_{u})$. The released magnetic energy
partitions into three distinct fractions: $(B_u+B_{z,u})/(2
B_u)$ for ion bulk flow energy associated with $V_{ix}$;
$(B_u-B_{z,u})(2B_{z,u}-B_u)/(2B_u^2)$ for ion bulk flow energy in
$V_{iz}$; and $(B_u-B_{z,u})^2/B_u^2$ for ion thermal energy. The fractions total to unity. They can be tested by the same set of simulations used
in Fig.~\ref{fig13}. In the exhaust, we calculate the ratio of these components of the ion energy (normalized by the number of ions) to the available magnetic
energy per particle $m_iC_{Ax,u}^2/(1+B_{z,u}/B_{u})$ and we plot them as
a function of $m_iC_{Ax,u}^2/(1+B_{z,u}/B_{u})$ in
Fig.~\ref{fig15}. The summation of the fraction of all forms is close to unity
at low-$\beta$, suggesting that our prediction of the
available magnetic energy per particle is correct.  Each line approaches
a constant and agrees reasonably well with the corresponding
predicted partition by anisotropic MHD in the low initial $\beta$
limit plotted in red.

\section{Conclusion}
In this paper we report the results of low-$\beta$ guide field particle-in-cell Riemann simulations with high ion-electron mass ratio to explore the particle heating in reconnection outflows far downstream from the x-line. Comparison with conventional reconnection simulations shows that Riemann simulations can produce comparable results when the simulation parameters overlap. Thus, Riemann simulations are good proxies of reconnection simulations and can be useful to explore the low-$\beta$ regime with more realistic parameters than is possible with full 2D reconnection simulations.

The results of Riemann simulations in the low-$\beta$ regime show that the RDs and SSs associated with reconnection clearly separate from one another, steadily moving outwards from the exhaust midplane. The steady expansion of the exhaust, as long as the domain is large enough, should continue without bound, suggesting that particle heating in the exhausts can extend to macroscopic scales in the
corona.
There is ion and electron heating between two SSs and electron heating between the RD and SS. The latter produces shoulders in the electron temperature profile that extend across the entire exhaust.
The heating mechanisms downstream of the SSs differ from those between
the RD and SS. Ions are accelerated by the RD magnetic field tension
and gain bulk flow energy along the $x$ direction (the reconnection
exhaust) and in the out-of-plane $z$ direction.  Electrons are
controlled by the electric potential that forms to produce the
localized parallel current to support the magnetic rotation at the RDs
and to maintain zero current elsewhere. These potentials partially
confine electrons within the exhaust.  The two RDs, however, have
different confinement characteristics. A higher fraction of electrons
are confined by the RD where the electric field driving the current at
the RD acts in the same direction as a confining electric field.  The
ion beams produced at the RDs counterstream across the midplane of the
exhaust and create a region of high density (a factor of two above the
upstream density) that defines the domain between the SSs. The
increase of the ion density leads to a region of high potential
between the SSs to confine downstream electrons to maintain charge
neutrality. The potential accelerates electrons from upstream of the
SSs towards downstream and traps a fraction of them, modestly
increasing the downstream electron temperature.

The heating of ions and electrons as a function of available magnetic
energy per particle reveals distinct differences between the two
species. The ion heating exhibits a roughly linear scaling with
available magnetic energy while the electron heating reaches a plateau
in the low-$\beta$ limit. The consequence is that the electron energy
increment is only of the same order as the upstream temperature. This
is in contrast to the linear scaling for both ions and electrons
that would be expected if the heating were simply proportional to the
available magnetic energy per particle\cite{Phan2013,Phan2014}. The
special scaling for electrons originates from the quasineutrality
requirement, which prohibits strong electron heating even with large
available magnetic energy per particle. As a result of this scaling,
ion heating dominates over electron heating in the low-$\beta$ limit
and the energy partition reduces to an anisotropic MHD prediction with
electron energy gain neglected.

Rowan et al.\cite{Rowan2019} have also investigated guide field reconnection heating
and energy partition with realistic mass ratio and low $\beta$. They concluded that electrons rather than ions gained most of the released energy in the strong guide field limit. However, they explored the trans-relativistic regime with magnetization
$\sigma=B^2/4\pi n m_i c^2\sim0.1$. This translates to an electron Alfv\'en speed close to $c$. Around the x-line and along magnetic separatrices the electron velocity approaches the electron Alfv\'en speed so electrons can approach relativistic velocities in a single x-line encounter. In the non-relativistic regime under consideration here, in which most electrons bypass the x-line and enter the exhaust downstream, the electrons gain negligible energy in a single passage through the exhaust. As a consequence, it is the ions rather than electrons that gain significant energy in a single interaction with the rotational discontinuity that bounds the reconnection exhaust. The ions therefore gain the most energy in the non-relativistic limit. 

The fundamental physics revealed in this study has broad implications to the inner heliosphere and the corona where reconnection plays a role in magnetic energy conversion.
This study specifically raises questions about how electrons gain
significant energy in the single x-line model of reconnection-driven
flare energy release. With the electron energy gain controlled by
potentials in our picture, neither very energetic electron nor very strong electron heating can take place in single x-line reconnection exhausts. The conventional picture of strong electron heating at the slow shocks produced during reconnection \cite{Tsuneta1996,Longcope2010} therefore fails. Further, the generation of an energetic electron  powerlaw tail up to energies of the order of an MeV as observed in large solar flares\cite{RPLin2003}, is also not possible in a single exhaust.
This study suggests that other mechanisms are required to explain electron energy gain in solar flares, such as multiple x-line reconnection \cite{Drake2006,Drake2013,Dahlin2015,Dahlin2017}.

\begin{acknowledgments}
Qile Zhang acknowledges helpful conversations with Colby C. Haggerty.
This work was supported by NSF Grant Nos. PHY1805829 and PHY1500460,
NASA Grant Nos. NNX14AC78G and NNX17AG27G, and the FIELDS team of the
Parker Solar Probe (NASA Contract No. NNN06AA01C).  Simulations were
carried out at the National Energy Research Scientific Computing
Center. Simulation data are available on request.
\end{acknowledgments}

\appendix*
\section{Calculation of anisotropic MHD solution}
Since we are looking at symmetric reconnection, we only need to consider one side of the domain with one RD and one SS. According to Lin et al.\cite{Lin1993}, with pressure anisotropy, the Rankine-Hugoniot jump conditions of each discontinuity (RD or SS) are:
\begin{equation}
\begin{split}
&[\rho V_y]=0\\
&[\rho V_y \textbf{V\textsubscript{t}} - \frac{B_y \textbf{B\textsubscript{t}}}{4 \pi}+\frac{B_y \textbf{B\textsubscript{t}}}{8 \pi}(\beta_\parallel-\beta_\perp)]=0\\
&[\rho V_y^2+P_\perp+\frac{B^2}{8 \pi}+\frac{B_y^2}{8 \pi}(\beta_\parallel-\beta_\perp)]=0\\
&[\left(\frac{1}{2}\rho V^2+\frac{5}{2}P+\frac{B^2}{4 \pi}-\frac{B^2}{24 \pi}(\beta_\parallel-\beta_\perp)\right)V_y-\left(1-\frac{1}{2}(\beta_\parallel-\beta_\perp)\right)\frac{B_y \textbf{B\textsubscript{t}}}{4 \pi}\cdot \textbf{V\textsubscript{t}}\\
	&-\left(1-\frac{1}{2}(\beta_\parallel-\beta_\perp)\right)\frac{B_y^2}{4 \pi}V_y]=0\\
&[B_y\textbf{V\textsubscript{t}}-V_y\textbf{B\textsubscript{t}}]=0
\end{split}
\end{equation}
,where $\rho=n m_i$, $\beta_\parallel$ and $\beta_\perp$ are plasma beta parallel and perpendicular to the local field, and $P=(P_\parallel+2 P_\perp)/3$. Subscript "t" means tangential to the shock surface. 

In the low initial $\beta$ limit, the perpendicular temperature (and thus pressure) throughout the solution can be neglected due to the conservation of magnetic moment. Also the parallel pressure upstream of the RD can be neglected. Similar to Liu et al.\cite{Liu2012}, since $B_y \ll B$ and $V_y$ is of the order of $C_{Ay}$ in reconnection, to the lowest order the equations can be simplified to the following:
\begin{equation}
\begin{split}
&[\rho V_y]=0\\
&[\rho V_y \textbf{V\textsubscript{t}} - \frac{B_y \textbf{B\textsubscript{t}}}{4 \pi}(1-\frac{\beta_\parallel}{2})]=0\\
&[\frac{B_t^2}{8 \pi}]=0\\
&[\left(\frac{1}{2}\rho \textbf{V\textsubscript{t}}^2+\frac{5}{2}3P_\parallel+\frac{B_t^2}{4 \pi}-\frac{B_t^2}{24 \pi}\beta_\parallel\right)V_y-\left(1-\frac{1}{2}\beta_\parallel\right)\frac{B_y \textbf{B\textsubscript{t}}}{4 \pi}\cdot \textbf{V\textsubscript{t}}]=0\\
&[B_y\textbf{V\textsubscript{t}}-V_y\textbf{B\textsubscript{t}}]=0
\end{split}
\end{equation}
There are a total of 7 jump conditions and 7 downstream unknowns here applicable to each RD or SS. The unknowns are $B_x, B_z, V_x, V_y, V_z, n,T_\parallel$. Note that the jump conditions for the RD, under the low-$\beta$ assumption, will reduce to that of isotropic MHD. In addition, there are three more unknowns: the speeds in y direction in the lab frame of the SS, RD and the plasma upstream of the RD. Note that $V_x$ and $V_z$ upstream of the RD are zero in the lab frame. Since the reconnection is symmetric here, we also have three more constraint equations for the quantities downstream of SS, which are $B_x=0, V_y=0, V_z=0$ in the lab frame.  So with the same number of equations as the unknowns, we can obtain a solution for all these physical quantities. 

In the solution, the speeds in y direction in the lab frame of the SS, RD and the plasma upstream of RD are $-sV_0 B_{y,u}/B_u$, $-s(C_{Ay,u}-V_0 B_{y,u}/B_u)$ and $sV_0 B_{y,u}/B_u$. 
Other quantities are: between the RD and SS, $B_x=0, B_z=B_u, n=n_0, V_{x}=-sB_{x,u}/\sqrt{4 \pi n_0 m_i},
V_{z}=s(B_{u}-B_{z,u})/\sqrt{4 \pi n_0 m_i}, T_\parallel=0, T_\perp=0$. Between the two SSs, $B_x=0, B_z=B_u, n=2n_0, V_{x}=-sB_{x,u}/\sqrt{4 \pi n_0 m_i},
V_{z}=0, T_\parallel=((B_{u}-B_{z,u})/\sqrt{4 \pi n_0 m_i})^2, T_\perp=0$.

There are a few notable features of this solution. In the lab frame there is no ${\bf E}\times {\bf B}$ flow in y-z plane between RD and SS as well as downstream of SS, so the field lines in the exhaust are simply stationary in y-z plane. The speed of the slow shock, if converted to a speed along the magnetic field, is about $V_0$, which is the same as the inflowing speed along the fields upstream of SS. 
The plasma upstream of the RD has an nonzero incoming speed and the RD is traveling with upstream $C_{Ay}$ relative to the upstream plasma.
\bibliography{reference}
\begin{widetext}
\begin{table}[ht]
\caption{Simulation parameters} % title of Table
\centering % used for centering table
\begin{tabular}{c c c c c c c c c c} % centered columns  (4 columns)
\hline\hline %inserts double horizontal lines
Run & $m_i/m_e$ & $B_{x,a}$ & $B_{z,a}$ & $T_i=T_e$ &  dims & $L_y \times L_x \times L_z$ &$c^2$&dx&dt\\ [0.5ex] % inserts table
%heading
\hline % inserts single horizontal line
1 & 25 & 1 & 1 & 0.05&2&102.4$\times$409.6 $\times$ 0&$45$&0.0125&5.9e-3 \\ % inserting body of the table
2 & 25 & 1 & 1 & 0.05&1&102.4$\times$0.2$\times$0&$45$&0.0125&5.9e-3 \\
3 & 400 & 1 & 2 & 0.02&2&22.9$\times$0$\times$0.54 &$720$&0.0007&2e-4\\
4 & 100 & 1 & 2 & 0.02&2&22.9$\times$0$\times$1.08 &$180$&0.0014&6e-4\\
5 & 400 & 1 & 2 & 0.02&1&22.9$\times$0$\times$0.011 &$720$&0.0007&2e-4\\
6 & 400 & $1/\sqrt{20}$ & $1/\sqrt{20}$ & 0.005&1&45.8$\times$0$\times$0.0022&$720$&0.00056&1.56e-4\\
7 & 400 & $\sqrt{0.125}$ & $\sqrt{0.125}$ & 0.005&1&45.8$\times$0$\times$0.0022&$720$&0.00056&1.56e-4 \\
8 & 400 & $\sqrt{0.25}$ & $\sqrt{0.25}$ & 0.005&1&45.8$\times$0$\times$0.0022&$720$&0.00056&1.56e-4\\
9 & 400 & $\sqrt{0.5}$ & $\sqrt{0.5}$ & 0.005&1&45.8$\times$0$\times$0.0022 &$720$&0.00056&1.66e-4\\
10 & 400 & $\sqrt{0.75}$ & $\sqrt{0.75}$ & 0.005&1&45.8$\times$0$\times$0.0022&$720$&0.00056&1.56e-4 \\
11 & 400 & 1 & 1 & 0.005&1&45.8$\times$0$\times$0.00186&$720$&0.000466&1.3e-4\\ % [1ex] adds vertical space
12 & 800 & $\sqrt{2}$ & $\sqrt{2}$ & 0.005&1&45.8$\times$0$\times$0.0011&$1440$&0.00028&5.5e-5\\
13 & 1600 & 1 & 2 & 0.02&1&22.9$\times$0$\times$0.00447 &$2880$&0.00028&4e-5\\
14 & 400 & 1 & 2 & 0.02&1&45.8$\times$0$\times$0.011 &$720$&0.0007&2e-4\\[1ex] % [1ex] adds vertical space
\hline %inserts single line
\end{tabular}
\label{table1} % is used to refer this table in the text
\end{table}
\end{widetext}



\end{document}